\documentstyle[preprint,aps]{revtex}

\tightenlines

\begin{document}

\draft

\title{
Flavour symmetry breaking,
baryons magnetic moments,\\
and low energy phenomenology of hadrons
}

\author{
A.~Garc\'{\i}a
}
\address{
Departamento de F\'{\i}sica\\
Centro de Investigaci\'on y de Estudios Avanzados del IPN\\
A.P. 14-740, C.P. 07000, M\'exico, D.F., MEXICO\\
}
\author{
R.~Huerta
and
G.~S\'anchez-Col\'on
}
\address{
Departamento de F\'{\i}sica Aplicada\\
Centro de Investigaci\'on y de Estudios Avanzados del IPN.
Unidad M\'erida\\
A.P. 73, Cordemex, C.P. 97310, M\'erida, Yucat\'an, MEXICO\\
}

\date{March, 2000}

\maketitle

\begin{abstract}
A priori mixings of eigenstates in physical states are quantum mechanical
effects well known in several realms of physics.
The possibility that such effects are also present in particle physics, in the
form of mixings that break flavor and parity symmetries, is studied.
Applications to non-leptonic and weak radiative decays of hyperons are
discussed.
The results are very encouraging but should be improved by eventually including the $W$ and $Z$ contributions, assumed small (non-enhanced) in this work.
\end{abstract}

\section{Introduction}
\label{introduction}

Because parity and strong flavors (strangeness, charm, etc.) symmetries are violated in nature, the physical (mass eigenstates) hadrons cannot be either parity or flavor eigenstates, i.e., the former must be admixtures of the latter.
It is generally believed that the breaking of flavor global groups is caused by
the mass differences of hadrons, but in such a way that parity and all flavors
are conserved, i.e., the mass operator of hadrons giving rise to such breakings
does not contain a piece that violates parity and flavor.
The flavor and parity mixings in physical hadrons are attributed to the
perturbative intervention of $W^{\pm}_{\mu}$ and $Z^0_{\mu}$ (parity mixing
only).
Precisely because such intervention is perturbative, such mixings can
appear only in higher orders of perturbation theory; thus, such mixings appear,
so to speak, {\em a posteriori.}

However, the possibility that the mass operator of hadrons does contain a
(necessarily) very small piece that is flavor and parity violating is not
excluded by any fundamental principle.
If such a piece does exist, then, the parity and flavor admixtures in hadrons
must come {\em a priori}, in a non-perturbative way.
It is not idle to emphasize that such a piece could not be attributed to the
$W^{\pm}_{\mu}$ and $Z^0_{\mu}$.

Our purpose in this work is to apply the a priori mixings scheme to non-leptonic and weak radiative decays of hyperons.

\section{A priori mixed hadrons}
\label{ansatz}

The implementation of a priori mixings for practical applications cannot, as
of today, be achieved from first principles, i.e., by starting from a model at
the quark level and then performing the QCD calculations to obtain the
physical hadrons and their couplings.
In order to proceed we must elaborate an ansatz.
We refer the reader to Ref.~\cite{jpg1} for a complete and detailed description of this ansatz.
Here, we only reproduce the expressions for the a priori mixed hadrons obtained with this ansatz (we shall restrict what follows to spin-1/2 baryons and spin-0 mesons):

\[
p_{ph} =
p_s -
\sigma\Sigma^+_s -
\delta\Sigma^+_p
+ \cdots,
\qquad
\Sigma^+_{ph} =
\Sigma^+_s +
\sigma p_s -
\delta' p_p
+ \cdots
\]

\[
\Sigma^-_{ph} =
\Sigma^-_s +
\sigma\Xi^-_s +
\delta\Xi^-_p
+ \cdots,
\qquad
\Xi^-_{ph} =
\Xi^-_s -
\sigma\Sigma^-_s +
\delta'\Sigma^-_p
+ \cdots
\]

\[
n_{ph} =
n_s +
\sigma(\frac{1}{\sqrt 2}\Sigma^0_s +
            \sqrt{\frac{3}{2}}\Lambda_s) +
\delta(\frac{1}{\sqrt 2}\Sigma^0_p +
            \sqrt{\frac{3}{2}}\Lambda_p)
+ \cdots
\]

\[
\Lambda_{ph} =
\Lambda_s +
\sigma\sqrt{\frac{3}{2}}(\Xi^0_s - n_s) +
\delta\sqrt{\frac{3}{2}}\Xi^0_p +
\delta'\sqrt{\frac{3}{2}}n_p
+ \cdots
\]

\[
\Sigma^0_{ph} =
\Sigma^0_s +
\sigma\frac{1}{\sqrt 2}(\Xi^0_s - n_s) +
\delta\frac{1}{\sqrt 2}\Xi^0_p +
\delta'\frac{1}{\sqrt 2}n_p
+ \cdots
\]

\[
\Xi^0_{ph} =
\Xi^0_s -
\sigma
(\frac{1}{\sqrt 2}\Sigma^0_s + \sqrt{\frac{3}{2}}\Lambda_s) +
\delta'
(\frac{1}{\sqrt 2}\Sigma^0_p + \sqrt{\frac{3}{2}}\Lambda_p)
+ \cdots
\]

\[
K^+_{ph} =
K^+_{p} - \sigma \pi^+_{p} - \delta' \pi^+_{s} + \cdots,
\qquad
K^0_{ph} =
K^0_{p} +
\sigma\frac{1}{\sqrt 2}\pi^0_{p} +
\delta'\frac{1}{\sqrt 2}\pi^0_{s} + \cdots
\]

\[
\pi^+_{ph} =
\pi^+_{p} + \sigma K^+_{p} - \delta K^+_{s} + \cdots
\]

\[
\pi^0_{ph} =
\pi^0_{p} -
\sigma\frac{1}{\sqrt 2}( K^0_{p} + \bar K^0_{p} ) +
\delta\frac{1}{\sqrt 2}( K^0_{s}- \bar K^0_{s} ) + \cdots
\]

\[
\pi^-_{ph} =
\pi^-_{p} + \sigma K^-_{p} + \delta K^-_{s} + \cdots
\]

\begin{equation}
\bar K^0_{ph} =
\bar K^0_{p} + \sigma \frac{1}{\sqrt 2}\pi^0_{p} -
\delta'\frac{1}{\sqrt 2}\pi^0_{s} + \cdots,
\qquad
K^-_{ph} =
K^-_{p} - \sigma \pi^-_{p} + \delta' \pi^-_{s} + \cdots
\label{cinco}
\end{equation}

\noindent
In these expressions the subindices $s$, and $p$ indicate
positive, and negative parity eigenstates and each physical
hadron is the mass eigenstate already observed.
We must point out that the previous mixings have a parallelism
at the quark level so that they should be necessary to develop a formulation
at that level.
This particular matter will not be tried here~\cite{rmf99}.

\section{Application to Non-Leptonic Decays}
\label{application2}

If strong-flavor and parity violating pieces in the mass operator of hadrons
exist they would lead to non-perturbative a priori mixings  of flavor and
parity eigenstates in physical (mass eigenstates) hadrons.
Then, two paths for weak decays of hadrons to occur would be open: the ordinary
one mediated by $W^\pm_\mu$ ($Z_{\mu}$) and a new one via the strong-flavor and
parity conserving interaction hamiltonians.
The enhancement phenomenon observed in non-leptonic decays of hyperons (NLDH)
could then be attributed to this new mechanism.
However, for this to be the case it will be necessary that a priori mixings
produce the well established predictions of the $|\Delta I|=1/2$
rule~\cite{marshak,donoghue}.

The a priori mixed hadrons will lead to NLDH via the parity and flavor
conserving strong interaction (Yukawa) hamiltonian $H_Y$.
The transition amplitudes will be given by the matrix elements
$\langle B_{ph}M_{ph}|H_Y|A_{ph}\rangle$, where $A_{ph}$ and $B_{ph}$ are the
initial and final hyperons and $M_{ph}$ is the emitted meson.
Using the above mixings, Eqs.~(\ref{cinco}), these amplitudes will have the
form $\bar{u}_B(A-B\gamma_5)u_A$, where $u_A$ and $u_B$ are four-component
Dirac spinors and the amplitudes $A$ and $B$ correspond to the parity
violating and the parity conserving amplitudes of the $W^{\pm}_{\mu}$ mediated
NLDH, although with a priori mixings these amplitudes are both actually parity
and flavor conserving.
As a first approximation we shall neglect isospin violations, i.e., we shall
assume that $H_Y$ is an $SU_2$ scalar.
However, we shall not neglect $SU_3$ breaking.
One obtains for $A$ and $B$ the results:

\[
A_1
=
\delta'
\sqrt 3 g^{{}^{p,sp}}_{{}_{p,p\pi^0}} +
\delta
(
g^{{}^{s,ss}}_{{}_{\Lambda,pK^-}} - g^{{}^{s,pp}}_{{}_{\Lambda,\Sigma^+\pi^-}}
)
,
\quad
A_2
=
-
\frac{1}{\sqrt{2}}
[
\delta'
\sqrt 3 g^{{}^{p,sp}}_{{}_{p,p\pi^0}} +
\delta
(
g^{{}^{s,ss}}_{{}_{\Lambda,pK^-}} - g^{{}^{s,pp}}_{{}_{\Lambda,\Sigma^+\pi^-}}
)
]
,
\]

\[
A_3
=
\delta
(
\sqrt 2 g^{{}^{s,ss}}_{{}_{\Sigma^0,p K^-}} +
\sqrt{\frac{3}{2}} g^{{}^{s,pp}}_{{}_{\Sigma^+,\Lambda\pi^+}} +
\frac{1}{\sqrt{2}} g^{{}^{s,pp}}_{{}_{\Sigma^+,\Sigma^+\pi^0}}
)
,
\]

\[
A_4
=
-\delta'
\sqrt 2 g^{{}^{p,sp}}_{{}_{p,p\pi^0}} +
\delta
(
\sqrt{\frac{3}{2}} g^{{}^{s,pp}}_{{}_{\Sigma^+,\Lambda\pi^+}} -
\frac{1}{\sqrt{2}} g^{{}^{s,pp}}_{{}_{\Sigma^+,\Sigma^+\pi^0}}
)
,
\]

\[
A_5
=
-\delta'
g^{{}^{p,sp}}_{{}_{p,p\pi^0}} -
\delta
(
g^{{}^{s,ss}}_{{}_{\Sigma^0,pK^-}} +
g^{{}^{s,pp}}_{{}_{\Sigma^+,\Sigma^+\pi^0}}
)
,
\quad
A_6
=
\delta'
g^{{}^{p,sp}}_{{}_{\Sigma^+,\Lambda\pi^+}} +
\delta
(
g^{{}^{s,ss}}_{{}_{\Xi^-,\Lambda K^-}} +
\sqrt 3 g^{{}^{s,pp}}_{{}_{\Xi^0,\Xi^0\pi^0}}
)
,
\]

\begin{equation}
A_7
=
\frac{1}{\sqrt{2}}
[
\delta'
g^{{}^{p,sp}}_{{}_{\Sigma^+,\Lambda\pi^+}} +
\delta
(
g^{{}^{s,ss}}_{{}_{\Xi^-,\Lambda K^-}} +
\sqrt 3 g^{{}^{s,pp}}_{{}_{\Xi^0,\Xi^0\pi^0}}
)
]
,
\label{cuatronl}
\end{equation}

\noindent
and

\[
B_1
=
\sigma
(
- \sqrt 3 g_{{}_{p,p\pi^0}} +
g_{{}_{\Lambda,pK^-}} - g_{{}_{\Lambda,\Sigma^+\pi^-}}
)
,
\quad
B_2
=
-
\frac{1}{\sqrt{2}}
\sigma
(
- \sqrt 3 g_{{}_{p,p\pi^0}} +
g_{{}_{\Lambda,pK^-}} - g_{{}_{\Lambda,\Sigma^+\pi^-}}
)
,
\]

\[
B_3
=
\sigma
(
\sqrt 2 g_{{}_{\Sigma^0,p K^-}} +
\sqrt{\frac{3}{2}} g_{{}_{\Sigma^+,\Lambda\pi^+}} +
\frac{1}{\sqrt{2}} g_{{}_{\Sigma^+,\Sigma^+\pi^0}}
)
,
\]

\[
B_4
=
\sigma
(
\sqrt 2 g_{{}_{p,p\pi^0}} +
\sqrt{\frac{3}{2}} g_{{}_{\Sigma^+,\Lambda\pi^+}} -
\frac{1}{\sqrt{2}} g_{{}_{\Sigma^+,\Sigma^+\pi^0}}
)
,
\]

\[
B_5
=
\sigma
(
g_{{}_{p,p\pi^0}} -
g_{{}_{\Sigma^0,pK^-}} - g_{{}_{\Sigma^+,\Sigma^+\pi^0}}
)
,
\quad
B_6
=
\sigma
(
- g_{{}_{\Sigma^+,\Lambda\pi^+}} +
g_{{}_{\Xi^-,\Lambda K^-}} +
\sqrt 3 g_{{}_{\Xi^0,\Xi^0\pi^0}}
)
,
\]

\begin{equation}
B_7
=
\frac{1}{\sqrt{2}}
\sigma
(
- g_{{}_{\Sigma^+,\Lambda\pi^+}} +
g_{{}_{\Xi^-,\Lambda K^-}} +
\sqrt 3 g_{{}_{\Xi^0,\Xi^0\pi^0}}
)
.
\label{cinconl}
\end{equation}

\noindent
The subindices $1, \dots, 7$ correspond to
$\Lambda\rightarrow p\pi^-$,
$\Lambda\rightarrow n\pi^0$,
$\Sigma^-\rightarrow n\pi^-$,
$\Sigma^+\rightarrow n\pi^+$,
$\Sigma^+\rightarrow p\pi^0$,
$\Xi^-\rightarrow \Lambda\pi^-$,
and
$\Xi^0\rightarrow \Lambda\pi^0$,
respectively.
The $g$-constants in these equations are Yukawa coupling constants (YCC)
defined by the matrix elements of $H_Y$ between flavor and parity eigenstates,
for example, by
$\langle B_{0s} M_{0p} |H_Y|A_{0p}\rangle=g^{{}^{p,sp}}_{{}_{A,BM}}$.
We have omitted the upper indeces in the $g$'s of the $B$ amplitudes because
the states involved carry the normal intrinsic parities of hadrons.
In Eqs.~(\ref{cinconl}) we have used the $SU_2$ relations
$g_{{}_{p,p\pi^0}}=-g_{{}_{n,n\pi^0}}=g_{{}_{p,n\pi^+}}/{\sqrt 2}
=g_{{}_{n,p\pi^-}}/{\sqrt 2}$,
$g_{{}_{\Sigma^+,\Lambda\pi^+}}=g_{{}_{\Sigma^0,\Lambda\pi^0}}
=g_{{}_{\Sigma^-,\Lambda\pi^-}}$,
$g_{{}_{\Lambda,\Sigma^+\pi^-}}=g_{{}_{\Lambda,\Sigma^0\pi^0}}$,
$g_{{}_{\Sigma^+,\Sigma^+\pi^0}}=-g_{{}_{\Sigma^+,\Sigma^0\pi^+}}
=g_{{}_{\Sigma^-,\Sigma^0\pi^-}}$,
$g_{{}_{\Sigma^0,pK^-}}=g_{{}_{\Sigma^-,nK^-}}/{\sqrt 2}
=g_{{}_{\Sigma^+,p\bar K^0}}/{\sqrt 2}$,
$g_{{}_{\Lambda,pK^-}}=g_{{}_{\Lambda,n\bar K^0}}$,
$g_{{}_{\Xi^0,\Xi^0\pi^0}}=g_{{}_{\Xi^-,\Xi^0\pi^-}}/{\sqrt 2}$,
$g_{{}_{\Xi^-,\Lambda K^-}}=-g_{{}_{\Xi^0,\Lambda \bar K^0}}$,
and
$g_{{}_{\Lambda,\Lambda \pi^0}}=0$.
Similar relations are valid within each set of upper indeces, e.g.,
$g^{{}^{p,sp}}_{{}_{p,p\pi^0}}=-g^{{}^{p,sp}}_{{}_{n,n\pi^0}}$, etc.;
the reason for this is that mirror hadrons may
be expected to have the same strong-flavor assignments as ordinary hadrons.
Thus, for example, $\pi^+_{s} $, $\pi^0_{s} $, and $\pi^-_{s} $ form an
isospin triplet, although a different one from the ordinary $\pi^+_{p} $,
$\pi^0_{p} $, and $\pi^-_{p} $ isospin triplet.
These latter relations have been used in Eqs.~(\ref{cuatronl}).

From the above results one readily obtains the equalities:

\begin{equation}
A_2 =
-\frac{1}{\sqrt{2}} A_1,\ \ \ \ \ \
A_5 =
\frac{1}{\sqrt{2}}
(
A_4-A_3
),\ \ \ \ \ \
A_7 =
\frac{1}{\sqrt{2}} A_6,
\label{seisnl}
\end{equation}

\begin{equation}
B_2 =
-\frac{1}{\sqrt{2}} B_1,\ \ \ \ \ \
B_5 =
\frac{1}{\sqrt{2}}
(
B_4-B_3
),\ \ \ \ \ \
B_7 =
\frac{1}{\sqrt{2}} B_6.
\label{sietenl}
\end{equation}

\noindent
These are the predictions of the $|\Delta I|=1/2$ rule.
That is, a priori mixings in hadrons as introduced above lead to the
predictions of the $|\Delta I|=1/2$ rule, but notice that they do not lead to
the $|\Delta I|=1/2$ rule itself.
This rule originally refers to the isospin covariance properties of the
effective non-leptonic interaction hamiltonian to be sandwiched between
strong-flavor and parity eigenstates.
The $I=1/2$ part of this hamiltonian is enhanced over the $I=3/2$ part.
In contrast, in the case of a priori mixings $H_Y$ has been assumed to be
isospin invariant, i.e., in this case the rule should be called a
$\Delta I=0$ rule.

It must be stressed that the results~(\ref{seisnl}) and (\ref{sietenl}) are very
general: (i) the predictions of the $|\Delta I|=1/2$ rule are obtained
simultaneously for the $A$ and $B$ amplitudes, (ii) they are independent of
the mixing angles $\sigma$, $\delta$, and $\delta'$, and (iii) they are also
independent of particular values of the YCC.
They will be violated by isospin breaking corrections.
So, they should be quite accurate, as is experimentally the case.

A detailed comparison with all the experimental data available in these decays
requires more space and is presented separately\cite{nll}.
Nevertheless, we shall briefly mention a few very important results.

First, the experimental $B$ amplitudes (displayed in
Table~\ref{table1nl})
are reproduced within a few percent by accepting that the YCC are given by the
ones observed in strong interactions\cite{dumbrajs}, an assumption which
cannot be avoided in this approach.
The best predictions for these amplitudes are
$B_1=22.11\times 10^{-7}$,
$B_2=-15.63\times 10^{-7}$,
$B_3=1.39\times 10^{-7}$,
$B_4=-42.03\times 10^{-7}$,
$B_5=-30.67\times 10^{-7}$,
$B_6=17.45\times 10^{-7}$,
and
$B_7=12.34\times 10^{-7}$.
The only unknown parameter $\sigma$ is determined at
$(3.9\pm 1.3)\times 10^{-6}$.
We quote the experimental values of the $B$ amplitudes in the natural scale of
$10^{-7}$, see Ref.~\cite{donoghue}.
Their signs are free to choose; actually, the comparison with theoretical
predictions is only meaningful for their magnitudes.
The signs we display are for convenience only.
This is not the case for the signs in the $A$ amplitudes.

Second, although the $A$ amplitudes involve new YCC, an important prediction is
already made in Eqs.~(\ref{cuatronl}).
Once the signs of the $B$ amplitudes are fixed, one is free to fix the signs of
four $A$ amplitudes --- say, $A_1>0$, $A_3<0$, $A_4<0$, $A_6<0$ --- to match
the signs of the corresponding experimental $\alpha$ asymmetries, namely,
$\alpha_1>0$, $\alpha_3<0$, $\alpha_4>0$, $\alpha_6<0$.
Then the signs of $A_2<0$, $A_5>0$, and $A_7<0$ are fixed by
Eqs.~(\ref{cuatronl}) and the fact that $|A_4|\ll |A_3|$.
In turn the signs of the corresponding $\alpha$'s are fixed.
These three signs agree with the experimentally observed ones, namely,
$\alpha_2>0$, $\alpha_5<0$, $\alpha_7<0$.

A detailed comparison of the $A$ amplitudes with experiment is limited by our
current inability to compute well with QCD.
However, one may try simple and argumentable new assumptions to make
predictions for such amplitudes.
Since QCD has been assumed to be common to both ordinary and mirror quarks, it
is not unreasonable to expect that the magnitudes of the YCC in the $A$
amplitudes have the same magnitudes as their corresponding counterparts in the
ordinary YCC of the $B$ amplitudes.
The relative signs may differ, however.
Introducing this assumption we obtain the predictions for the $A$ amplitudes
displayed in Table~\ref{table1nl}.
The predictions for the $B$ amplitudes must also be redone, because
determining the $A$ amplitudes alone may introduce small variations in the YCC
that affect importantly the $B$ amplitudes, i.e., both the $A$ and $B$
amplitudes must be simultaneously determined, the $B$'s act then as extra
constraints on the determination of the $A$'s.
The new predictions for the $B$'s are also displayed in Table~\ref{table1nl}.
In obtaining Table~\ref{table1nl} we have actually used the experimental decay
rates $\Gamma$ and $\alpha$ and $\gamma$ asymmetries\cite{pdg}, but we only
display the experimental and theoretical amplitudes.

The predictions for the $A$'s agree very well with experiment to within a few
percent, while the predictions for the $B$'s remain as before.
The a priori mixing angles are determined to be
$|\delta|=(0.23\pm 0.07)\times 10^{-6}$,
$|\delta'|=(0.26\pm 0.07)\times 10^{-6}$,
and
$\sigma=(4.9\pm 1.5)\times 10^{-6}$.
This last value of $\sigma$ is consistent with the previous one.
The overall sign of the new YCC can be reversed and the new overall sign can be
absorbed into $\delta$ and $\delta'$.
This can be done partially in the group of such constants that accompanies
$\delta$ or in the group that accompanies $\delta'$ or in both.
Because of this, we have determined only the absolute values of $\delta$ and
$\delta'$.
In order to emphasize this fact we have inserted absolute value bars on
$\delta$ and $\delta'$.
The more detailed analysis of the comparison of the $A$'s and $B$'s with
experiment is presented in Ref.~\cite{nll} and it indicates that violations of the $|\Delta I| = 1/2$ rule affect the values of the a priori mixing angles and one should take more conservative ones as their estimates in NLDH, namely,
$\sigma = (4.9 \pm 2.0)\times 10^{-6}$,
$|\delta|= (0.22 \pm 0.09)\times 10^{-6}$,
and
$|\delta'| = (0.26 \pm 0.09)\times 10^{-6}$.

The above results, especially those of Eqs.~(\ref{seisnl}) and (\ref{sietenl})
and the determination of the amplitudes, satisfy some of the most important
requirements that a priori mixings must meet in order to be taken seriously as
an alternative to describe the enhancement phenomenon observed in non-leptonic
decays of hadrons.

\section{Application to Weak Radiative Decays}
\label{application}

In contrast to $W^{\pm}_{\mu}$ mediated
weak radiative decays, a priori mixed baryons can produce weak radiative
decays via the ordinary electromagnetic interaction hamiltonian
$H^{em}_{int}=eJ^{em}_{\mu}A^{\mu}$,
where $J^{em}_{\mu}$ is the familiar e.m. current operator which is a flavor
conserving Lorentz proper four-vector.
That is, a priori mixings in baryons lead to weak radiative decays that in
reality are ordinary parity and flavor conserving radiative decays, whose
transition amplitudes are non-zero only because physical baryons are not
flavor and parity eigenstates\cite{univ}.

The radiative decay amplitudes we want are given by the usual matrix elements
$\langle\gamma,B_{ph}|H^{em}_{int}|A_{ph}\rangle$, where
$A_{ph}$ and $B_{ph}$ stand for hyperons.
A very simple calculation leads to the following hadronic matrix elements

\[
\langle p_{ph} | J^{\mu}_{em} | \Sigma^+_{ph} \rangle =
\bar u_p [ \sigma ( f^{\Sigma^+}_2 - f^p_2 ) +
           (\delta'f^p_2 - \delta f^{\Sigma^+}_2) \gamma^5 ]
i\sigma^{\mu\nu}q_{\nu} u_{\Sigma^+}
\]

\[
\langle \Sigma^-_{ph} | J^{\mu}_{em} | \Xi^-_{ph} \rangle =
\bar u_{\Sigma^-} [ \sigma ( f^{\Xi^-}_2 - f^{\Sigma^-}_2 ) +
                    (\delta' f^{\Sigma^-}_2 - \delta f^{\Xi^-}_2) \gamma^5 ]
i\sigma^{\mu\nu}q_{\nu} u_{\Xi^-}
\]

\begin{eqnarray}
\langle n_{ph} | J^{\mu}_{em} | \Lambda_{ph} \rangle
& = &
\bar u_n
\left\{
\sigma \left[ \sqrt{\frac{3}{2}} ( f^{\Lambda}_2 - f^n_2 ) +
                   \frac{1}{\sqrt 2} f^{\Sigma^0\Lambda}_2 \right]
\right.
\nonumber \\
& &
\left.
+ \left[
\sqrt{\frac{3}{2}}
(\delta' f^n_2  - \delta f^{\Lambda}_2)
-
\delta \frac{1}{\sqrt 2} f^{\Sigma^0\Lambda}_2
\right]
\gamma^5
\right\}
i\sigma^{\mu\nu}q_{\nu} u_{\Lambda}
\nonumber
\end{eqnarray}

\begin{eqnarray}
\langle \Lambda_{ph} | J^{\mu}_{em} | \Xi^0_{ph} \rangle
& = &
\bar u_{\Lambda}
\left\{
\sigma \left[ \sqrt{\frac{3}{2}} ( f^{\Xi^0}_2 - f^{\Lambda}_2 ) -
         \frac{1}{\sqrt 2} f^{\Sigma^0\Lambda}_2 \right]
\right.
\nonumber \\
& &
\left.
+ \left[
\sqrt{\frac{3}{2}} (\delta' f^{\Lambda}_2 - \delta f^{\Xi^0}_2 )
+
\delta' \frac{1}{\sqrt 2} f^{\Sigma^0\Lambda}_2
\right]
\gamma^5
\right\}
i\sigma^{\mu\nu}q_{\nu} u_{\Xi^0}
\nonumber
\end{eqnarray}

\begin{eqnarray}
\langle \Sigma^0_{ph} | J^{\mu}_{em} | \Xi^0_{ph} \rangle
& = &
\bar u_{\Sigma^0}
\left\{
\sigma \left[ \frac{1}{\sqrt 2} ( f^{\Xi^0}_2 - f^{\Sigma^0}_2 ) -
         \sqrt{\frac{3}{2}} f^{\Sigma^0\Lambda}_2 \right]
\right.
\nonumber \\
& &
\left.
+ \left[
\frac{1}{\sqrt 2} (\delta' f^{\Sigma^0}_2 - \delta f^{\Xi^0}_2 )
+
\delta' \sqrt{\frac{3}{2}}f^{\Sigma^0\Lambda}_2
\right]
\gamma^5
\right\}
i\sigma^{\mu\nu}q_{\nu} u_{\Xi^0}
\label{seis}
\end{eqnarray}

\noindent
In these amplitudes only contributions to first order in $\sigma$, $\delta$,
and $\delta'$ need be kept.
Each matrix element is flavor and parity conserving and can be expanded in
terms of charge $f_1(0)$ form factors and anomalous magnetic $f_2(0)$ form
factors.
Because the charges of the positive- and negative-parity parts of the same
physical wave function are equal and such charges are controlled by the
generator property of $J_\mu$ all the $f_1$'s cancel away and only the $f_2$
contribute.
The $f_2$ between $s$ and $p$ parts and between $p$ and $s$ parts can be
identified with the $f_2$ between $s$ and $s$ parts, provided that a relative
minus sign be present between the former two in order to respect hermicity.
Notice that the amplitudes~(\ref{seis}) are all of the form
$\bar{u}_B(C+D\gamma_5)i\sigma^{\mu\nu}q_{\nu}u_A$, where $C$ is the so-called
parity conserving amplitude and $D$ is the so-called parity violating one.
We stress, however, that in this model both $C$ and $D$ are parity and flavor
conserving.

We shall compare Eqs.~(\ref{seis}) with experiment, ignoring the contributions
of $W^{\pm}_{\mu}$ amplitudes.
We shall do this in order to be able to appreciate to what extent a priori
mixings provide on their own right a framework to describe weak radiative
decays.

In principle, we have information about all the quantities that appear in
these amplitudes for WRDH.
The a priori angles are known from NLDH and the $f_2$ can be related to the
measured total magnetic moments of spin 1/2 baryons.
The latter values are displayed in the second column of
Table~\ref{tablai}~\cite{pdg}.
However, it is important that the mixing angles be determined independently in
WRDH and, accordingly, we shall use them as free parameters in the remaining
of this paper.
How the $f_2$'s are related to the observed total magnetic moments is a
question we shall deal with in steps.
As a first approximation we shall assume that the $f_2$'s are related to the
$\mu$'s of Table~\ref{tablai} by the formula $\mu^{exp}_A = e_A + f^A_2$ where
$A$ is a baryon and $e_A$ its charge.
Thus, for example, $f^p_2$ obeys the relationship $\mu^{exp}_p = 1 + f^p_2$
(in nuclear magnetons), etc. Using this assumption we may compare with the
experimental data of WRDH.
The predictions obtained are displayed in the columns~I of
Table~\ref{tablai}.

These first results are not quite good yet but they have a qualitative value.
The main point is that the a priori mixing angles come out with the same order
of magnitude observed in NLDH, which is very encouraging.
The predictions for the observables are some very good, some good, but some
show important deviations.
The latter still have qualitative value, but should be improved.
The  values of the $\mu$'s agree fairly well with their experimental
counterparts.

As an intermediate step in this analysis it turns out to be very helpful to
see what are the values of the total magnetic moments required to reproduce
well the experimental observables of WRDH.
This is achieved by relaxing the error bars of the measured $\mu$'s up to
10\% of the corresponding central values and, then, repeating the previous
step.
The results are displayed in the columns~II of Table~\ref{tablai}.
The experimental data are very well reproduced now, but at the expense of
sizable (several percent) changes in the $\mu$'s and new values for the mixing
angles.

This second step clearly shows that we must accept that our first
approximation ---that of identifying the experimentally measured $\mu$'s with
the ones that are actually related to the $f_2$'s in this approach to
WRDH--- must be improved.
The $\mu$'s to be used for determining the $f_2$'s in the WRDH amplitudes are
really transition magnetic moments.
For example, the measured value of $\mu_p$ corresponds to the matrix element
$\langle p_{ph}|J^\mu_{em}|p_{hp}\rangle \simeq
\langle p_{os}|J^\mu_{em}|p_{os}\rangle$,
where both physical wave functions carry the mass $m_p$.
In contrast, the $\mu_p$ the appears in $\Sigma^+ \to p\gamma$ corresponds to
a matrix element whose bra carries the mass $m_p$ and whose ket carries the
mass $m_{\Sigma^+}$.
So, the normalization of $\mu_p$ originating in the matrix element
$\langle p_{ph}|J_\mu |\Sigma^+_{ph}\rangle$
should be related to both masses, $m_p$ and $m_{\Sigma^+}$.
It is in this sense that the magnetic moments that we must use are transition
magnetic moments.

The natural normalization of magnetic moments is determined by the Gordon
decomposition.
Using this expansion for guidance, then, for example, $\mu_p$ should be
normalized to $m_p + m_{\Sigma^+}$ and not to $2m_p$, etc.
One can see already a qualitative indication of this happening in the first
column~II in Table~\ref{tablai}, the changes in the $\mu$'s are
systematically in this direction.
$\mu_p$, $\mu_n$, $\mu_{\Xi^-}$, $\mu_{\Sigma^-}$,
$\mu_{\Sigma^+}$, $\mu_{\Xi^0}$, and $\mu_{\Sigma^0}$ appear to become
smaller or larger according to such changes in normalization.
$\mu_\Lambda$ and $\mu_{\Sigma^0\Lambda}$ are mixed cases because they appear
in two or three decays and will be required to be reduced or to be increased
in going from one case to another and, therefore, Table~\ref{tablai} cannot
provide a clear cut tendency.

Our third step is to improve our approximation following the above discussion.
One must change the normalization of the total magnetic moments either by
applying, for example, the factor $(m_p+m_p)/(m_p+m_{\Sigma^+})$ to the
experimental $\mu_p$ or the inverse factor to the theoretical $\mu_p$ related
to $f^p_2$.
Numerically, either way leads to the same result.
For definiteness, we choose the former.
The corrected experimental values are displayed in column five of
Table~\ref{tablai}.
Then, recalculating everything lead to the predictions of columns labeled~III
of Table~\ref{tablai}.
The values of the mixing angles appear in the bottom of the last column of
this table.

The overall agreement is greatly improved, the experimental data are well
produced while keeping the magnetic moments in very good agreement with their
experimental counterparts.
The only deviations that merit further discussion appear in $\Gamma_2$ and in
$\mu_{\Sigma^-}$ which are intimately related.
These deviations are probably due to another one of our approximations:
ignoring the contributions of $W_\mu$~\cite{bassaleck}.
The agreement already obtained is probably the best one can hope for if one
stays short of actually incorporating the contributions of $W_\mu$, as we
have.

\section{Conclusions}
\label{discussion}

If a priori mixings are present, then weak decays may go via the flavor and
parity conserving hamiltonians of strong and electromagnetic interactions.
That is, with these mixings there would exist another mechanism to produce
weak radiative, non-leptonic, and rare mode decays of hadrons, in addition to
the already existing mechanisms provided by the $W^{\pm}_{\mu}$ and $Z^0_{\mu}$
bosons.

We are now in a position to conclude our present analysis.
To extend the credibility of the a priori mixing scheme it was very important
to be able to describe WRDH.
As we have shown above this is achieved.
However, the most important and stringent test is that the mixing angles share
a universality-like property.
The values for them obtained independently in WRDH are in very good agreement with the absolute values obtained for them in NLDH.
It is the passing this universality-like test that lends the strongest
support to the possibility that the above scheme may serve a framework for
the systematic description of the enhancement phenomenon in weak decays of
hadrons.
Also, the contributions of $W^{\pm}_{\mu}$ should be included at some point at
a, for consistency, small level, say, by assuming that $|\Delta  I|=1/2$
amplitudes are of the same order of magnitude as the $|\Delta I|=3/2$
amplitudes.

\acknowledgments

This work was partially supported by CONACYT (M\'exico).

\begin{table}
\caption{
Predictions for the $A$ amplitudes, along with the accompanying predictions for
the $B$ amplitudes, obtained by assuming that the magnitudes of the YCC of
Eqs.~(2)
match their corresponding counterparts in
Eqs.~(3).
The values of the YCC are listed in Ref.~[6].
All amplitudes are given in units of $10^{-7}$.
}
\label{table1nl}
\begin{tabular}
{
r@{$\rightarrow$}l
r@{.}l@{\,$\pm$\,}r@{.}l
d
r@{.}l@{\,$\pm$\,}r@{.}l
d
}
\hline
\multicolumn{2}{c}{Decay} &
\multicolumn{4}{c}{$B_{\rm exp}$} &
$B_{\rm th}$ &
\multicolumn{4}{c}{$A_{\rm exp}$} &
$A_{\rm th}$
\\
\hline
$\Lambda$ & $p\pi^-$ &
$-$22 & 09 & 0 & 44 &
$-$22.38 &
$-$3 & 231 & 0 & 020 &
$-$3.262
\\
$\Lambda$& $n\pi^0$ &
15 & 89 & 1 & 01 &
15.83 &
2 & 374 & 0 & 027 &
2.307
\\
$\Sigma^-$ & $n\pi^-$ &
1 & 43 & 0 & 17 &
1.34 &
$-$4 & 269 & 0 & 014 &
$-$4.264
\\
$\Sigma^+$ & $n\pi^+$ &
$-$42 & 17 & 0 & 18 &
$-$42.09 &
$-$0 & 140 & 0 & 027 &
$-$0.152
\\
$\Sigma^+$ & $p\pi^0$ &
\multicolumn{4}{c}{$-26.86{\ }^{+\ 1.10}_{-\ 1.36}$} &
$-$30.72 &
\multicolumn{4}{r}{$3.247{\ }^{+\ 0.089}_{-\ 0.116}$} &
2.907
\\
$\Xi^-$ & $\Lambda\pi^-$ &
$-$17 & 47 & 0 & 50 &
$-$17.27 &
4 & 497 & 0 & 020 &
4.521
\\
$\Xi^0$ & $\Lambda\pi^0$ &
$-$12 & 29 & 0 & 70 &
$-$12.21 &
3 & 431 & 0 & 055 &
3.197
\\
\hline
\end{tabular}
\end{table}

\begin{table}
\squeezetable
\caption{
Experimental values of the observables in WRDH and of the magnetic moments
(m.\ m.) of hyperons along with the predictions of the three cases considered.
The indeces 1,2,...,5 on the observables
correspond, respectively, to the decays
$\Sigma^+\to p\gamma$, $\Xi^-\to\Sigma^-\gamma$, $\Lambda\to n\gamma$,
$\Xi^0\to\Lambda\gamma$, and $\Xi^0\to\Sigma^0\gamma$.
The numbers in parenthesis in the m.\ m.\ indicate the decay in which they appear.
The values of the a priori mixing angles in column eight come from NLDH.
The mixing angles are in 10$^{-6}$, the decay rates are in $10^6$~sec$^{-1}$,
and the m.\ m.\ are in nuclear magnetons.
The only m.\ m.\ that has not been measured is $\mu_{\Sigma^0}$.
We have taken for it its $SU(3)$ estimate with a 10\% error bar.
}
\label{tablai}
\begin{tabular}
{
l
r@{.}l@{$\pm$}r@{.}l
d
d
r@{.}l@{$\pm$}r@{.}l
d
c
r@{.}l@{$\pm$}r@{.}l
d
d
d
}
&
\multicolumn{6}{c}{Magnetic moments} &
\multicolumn{5}{c}{Transition m.\ m.} &
&
\multicolumn{7}{c}{Observables and angles}
\\
&
\multicolumn{4}{c}{Exp.} &
I &
II &
\multicolumn{4}{c}{Exp.} &
III &
&
\multicolumn{4}{c}{Exp.} &
I &
II &
III
\\
\tableline
$\mu_p (1)$ &
2 & \multicolumn{3}{l}{793} &
2.793 &
2.745 &
2 & \multicolumn{3}{l}{463} &
2.463 &
$\Gamma_1$ &
15 & 65 & 0 & 88 &
11.55 &
15.60 &
14.62
\\
$\mu_n (3)$ &
$-$1 & \multicolumn{3}{l}{913} &
$-$1.913 &
$-$1.654 &
$-$1 & \multicolumn{3}{l}{750} &
$-$1.750 &
$\Gamma_2$ &
0 & 77 & 0 & 14 &
1.37 &
0.81 &
1.30
\\
$\mu_{\Xi^-} (2)$ &
$-$0 & 651 & 0 & 003 &
$-$0.652 &
$-$0.747 &
$-$0 & 683 & 0 & 003 &
$-$0.685 &
$\Gamma_3$ &
6 & 65 & 0 & 57 &
7.13 &
6.67 &
6.16
\\
$\mu_{\Sigma^-} (2)$ &
$-$1 & 160 & 0 & 025 &
$-$1.018 &
$-$0.868 &
$-$1 & 103 & 0 & 024 &
$-$0.958 &
$\Gamma_4$ &
3 & 66 & 0 & 56 &
5.26 &
3.68 &
4.38
\\
$\mu_\Lambda (3)$ &
$-$0 & 613 & 0 & 004 &
$-$0.611 &
$-$0.553 &
$-$0 & 665 & 0 & 004 &
$-$0.665 &
$\Gamma_5$ &
12 & 1 & 1 & 4 &
5.23 &
11.8 &
10.13
\\
$\mu_\Lambda' (4)$ &
\multicolumn{4}{c}{} &
&
&
$-$0 & 563 & 0 & 004 &
$-$0.562 &
$\alpha_1$ &
$-$0 & 76 & 0 & 08 &
$-$0.78 &
$-$0.75 &
$-$0.88
\\
$\mu_{\Sigma^+} (1)$ &
2 & 458 & 0 & 010 &
2.430 &
2.553 &
2 & 748 & 0 & 011 &
2.763 &
$\alpha_2$ &
\multicolumn{4}{c}{} &
$-$0.09 &
0.56 &
0.20
\\
$\mu_{\Xi^0} (4)$ &
$-$1 & 250 & 0 & 014 &
$-$1.271 &
$-$1.502 &
$-$1 & 353 & 0 & 015 &
$-$1.357 &
$\alpha_3$ &
\multicolumn{4}{c}{} &
$-$0.79 &
$-$0.83 &
0.76
\\
$\mu_{\Xi^0}' (5)$ &
\multicolumn{4}{c}{} &
&
&
$-$1 & 311 & 0 & 015 &
$-$1.305 &
$\alpha_4$ &
0 & 40 & 0 & 40 &
$-$0.86 &
0.18 &
0.27
\\
$\mu_{\Sigma^0\Lambda} (3)$ &
$-$1 & 610 & 0 & 080 &
$-$1.597 &
$-$1.626 &
$-$1 & 808 & 0 & 090 &
$-$1.659 &
$\alpha_5$ &
0 & 20 & 0 & 32 &
$-$0.46 &
$-$0.02 &
0.77
\\
$\mu_{\Sigma^0\Lambda}' (4)$ &
\multicolumn{4}{c}{} &
&
&
$-$1 & 529 & 0 & 076 &
$-$1.426 &
$\sigma$ &
4 & 9 & 2 & 0 &
\multicolumn{1}{r}{0.70$\pm$0.03} &
\multicolumn{1}{r}{1.72$\pm$0.46} &
\multicolumn{1}{r}{0.91$\pm$0.06}
\\
$\mu_{\Sigma^0\Lambda}'' (5)$ &
\multicolumn{4}{c}{} &
&
&
$-$1 & 482 & 0 & 074 &
$-$1.617 &
$\delta$ &
$|$0 & 22$|$ & 0 & 09 &
\multicolumn{1}{r}{0.03$\pm$0.08} &
\multicolumn{1}{r}{$-$0.40$\pm$0.19} &
\multicolumn{1}{r}{$-$0.11$\pm$0.08}
\\
$\mu_{\Sigma^0} (5)$ &
0 & 649 & 0 & 065 &
0.499 &
0.624 &
0 & 617 & 0 & 062 &
0.500 &
$\delta'$ &
$|$0 & 26$|$ & 0 & 09 &
\multicolumn{1}{r}{0.10$\pm$0.07} &
\multicolumn{1}{r}{$-$0.28$\pm$0.18} &
\multicolumn{1}{r}{$-$0.25$\pm$0.08}
\\
\end{tabular}
\end{table}

\end{document}